\documentclass[usenatbib]{mnras}

\usepackage{newtxtext,newtxmath}
\usepackage[T1]{fontenc}

\DeclareRobustCommand{\VAN}[3]{#2}
\let\VANthebibliography\thebibliography
\def\thebibliography{\DeclareRobustCommand{\VAN}[3]{##3}\VANthebibliography}

\usepackage{graphicx}	% Including figure files
\usepackage{amsmath}	% Advanced maths commands
\usepackage{svg}

\usepackage{booktabs}
\usepackage{siunitx}
\usepackage{multirow}
\usepackage{float}
\usepackage{placeins}
\usepackage{subcaption}
\usepackage{cuted}
\usepackage{array}
\usepackage{xcolor}
\usepackage{booktabs}

\definecolor{softgreen}{RGB}{34, 139, 34} % Softer green color
\newcommand{\greenuparrow}{\textcolor{softgreen}{$\uparrow$}} % Added space and bolder arrow
\definecolor{softred}{rgb}{0.8, 0.1, 0.1}
\newcommand{\reddownarrow}{\textcolor{softred}{$\downarrow$}}

\newif\ifdownsampled
\newif\ifsvg
\title[AstroM$^3$: A self-supervised multimodal model for astronomy]{AstroM$^3$: A self-supervised multimodal model for astronomy}

\author[M. Rizhko et al.]{
M.\ Rizhko,$^{1}$\thanks{E-mail: mariia.rizhko@berkeley.edu}
J.\ S.\ Bloom,$^{1,2}$
\\
$^{1}$University of California, Berkeley, Department of Astronomy, Berkeley, CA, USA\\
$^{2}$Lawrence Berkeley National Laboratory, Berkeley, CA, USA\\
}

\pubyear{2024}

\begin{document}
\label{firstpage}
\pagerange{\pageref{firstpage}--\pageref{lastpage}}
\maketitle

\begin{abstract}

While machine-learned models are now routinely employed to facilitate astronomical inquiry, model inputs tend to be limited to a primary data source (namely images or time series) and, in the more advanced approaches, some metadata. Yet with the growing use of wide-field, multiplexed observational resources, individual sources of interest often have a broad range of observational modes available. Here  we construct an astronomical multimodal dataset and propose AstroM$^3$, a self-supervised pre-training approach that enables a model to learn from multiple modalities simultaneously. Specifically, we extend the CLIP (Contrastive Language-Image Pretraining) model to a trimodal setting, allowing the integration of time-series photometry data, spectra, and astrophysical metadata.  In a fine-tuning supervised setting, our results demonstrate that CLIP pre-training improves classification performance for time-series photometry, where accuracy increases from 84.6\% to 91.5\%. Furthermore, CLIP boosts classification accuracy by up to 12.6\% when the availability of labeled data is limited, showing the effectiveness of leveraging larger corpora of unlabeled data. In addition to fine-tuned classification, we can use the trained model in other downstream tasks that are not explicitly contemplated during the construction of the self-supervised model. In particular we show the efficacy of using the learned embeddings for misclassifications identification, similarity search, and anomaly detection. One surprising highlight is the "rediscovery" of Mira subtypes and two Rotational variable subclasses using manifold learning and dimension reduction algorithm. To our knowledge this is the first construction of an $n>2$ mode model in astronomy. Extensions to $n>3$ modes is naturally anticipated with this approach. 

\end{abstract}


%%%%%%%%%%%%%%%%%%%%%%%%%%%%%%%%%%%%%%%%%%%%%%%%%%

%%%%%%%%%%%%%%%%% BODY OF PAPER %%%%%%%%%%%%%%%%%%

\begin{figure*}
    \ifsvg
        \includesvg[width=\textwidth]{images/astroclip}
    \else
        \includegraphics[width=\textwidth]{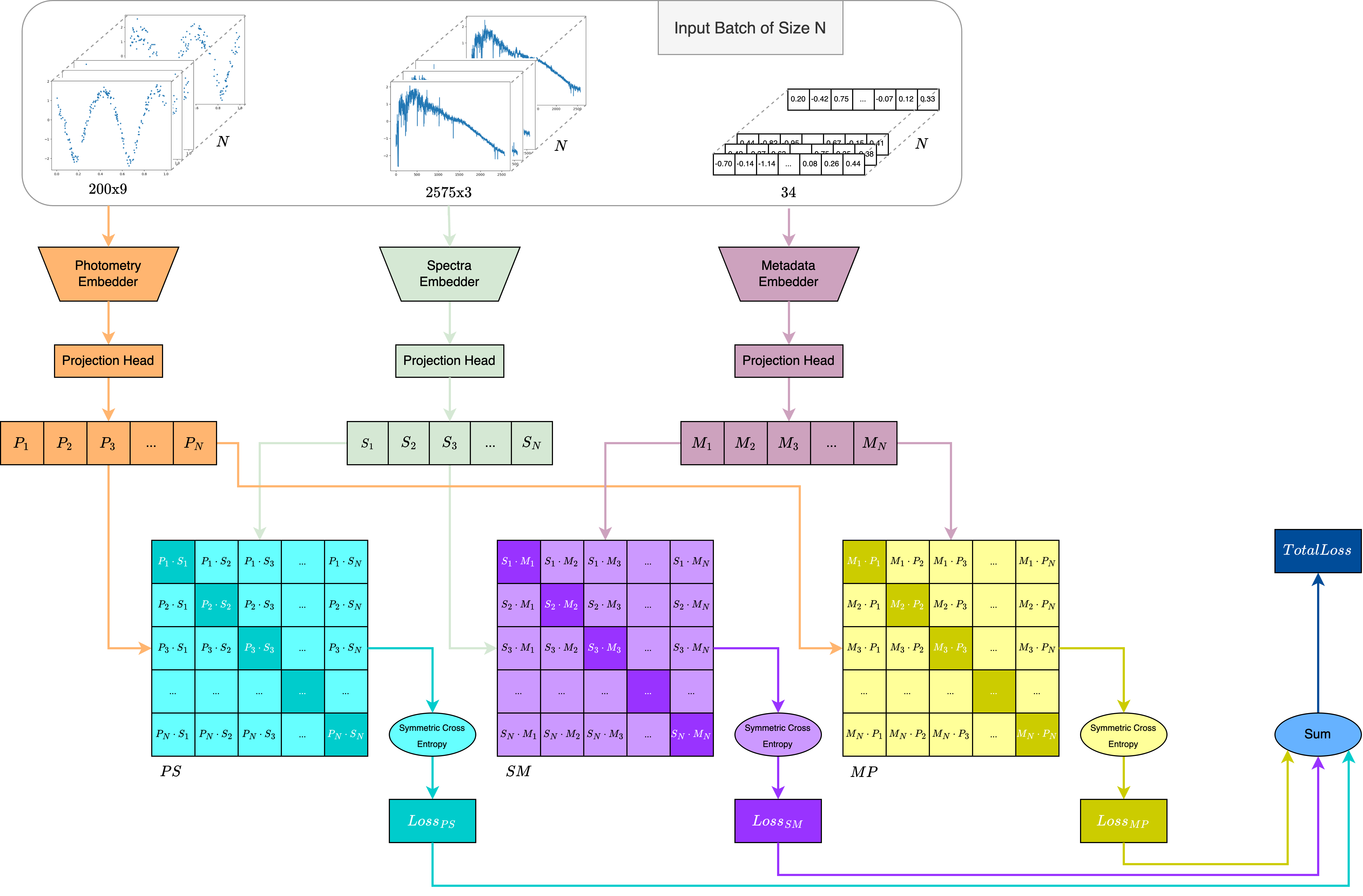}
    \fi
    \caption{Overview of the multimodal CLIP framework adapted for astronomy, incorporating three data modalities: photometric time-series, spectra, and metadata. Each modality is processed by a dedicated encoder to create embeddings, which are then mapped into a shared embedding space through projection heads. Pairwise similarity matrices align the embeddings across modalities, and a symmetric cross-entropy loss, computed over these matrices, optimizes the model. The total loss, derived from all pairwise losses, guides the model’s trimodal learning.}
    \label{fig:clip}
\end{figure*}

\section{Introduction}

Despite the vast volumes of publicly available raw astronomical data, with a few notable subfield exceptions, the application of machine learning to discovery and inference has yet to broadly permeate the field. One impediment stems from the challenge of fusing data across heterogeneous modes of  collection. Off-the-shelf architectures do not easily accommodate a mixture of irregularly sampled multi-spectral multi-scale heteroskedatic time-series data, images, spectra, and metadata. Another issue, arising in the classification context, is that very few ground-truth labels exist. This ``small label'' problem arose, for example, in \citet{2012ApJS..203...32R}, who sought to probabilistically classify 50,124 variable stars using only 810 labels over 28 classes. Last, models learned on a dataset from one survey do not easily transfer to other data collected on the same objects from different surveys (e.g., \citealt{2012PASP..124..280L,2021A&A...653A..22K}). Our self-supervised multimodal architecture addresses the first two challenges, establishing methods and milestones for a more generalized foundation model applicable to inference tasks on unseen survey data.

Our work builds upon the Contrastive Language-Image Pretraining (CLIP) framework, originally introduced by \cite{radford2021learning}; CLIP demonstrated the power of contrastive learning on large-scale image and text datasets to learn joint representations. Since its introduction, CLIP has been extensively researched and improved in various ways. For example, \cite{li2021supervision} enhanced data efficiency through supervision, while \cite{yao2021filip} focused on improving semantic alignment. \cite{cherti2023reproducible} introduced scaling laws, and \cite{sun2023eva} optimized the model for faster training. Additionally, CLIP has been combined with other pretraining objectives: \cite{mu2022slip} incorporated image self-supervision, and \cite{singh2022flava} along with \cite{li2022blip} added masked multimodal, image, and language modeling. Furthermore, CLIP has been extended to other modalities: audio-text  \citep{wu2023large},  video-text \citep{luo2021clip4clip, xu2021videoclip, ma2022x}, and point cloud-text \citep{zhang2022pointclip}. In the astronomical context, \cite{Parker_2024} used dual-mode CLIP on static-sky galaxy images and spectra. Closest to the approach of our work outside of astronomy, \cite{guzhov2022audioclip} adapted CLIP for use with three modalities: audio, image, and text. Given the proven versatility and success of CLIP in different domains, we build upon it herein. We extend CLIP to work on three modalities: time-series photometry, spectra, and metadata (see Figure \ref{fig:clip}). Our work, and a recent preprint from \citet{2024arXiv240816829Z}, are the first efforts to incorporate time-series data with CLIP, and our three-mode model represents a critical step towards the development of a foundational multimodal model for time-domain astronomy.

\section{Related Work}

Early classification-focused research used hand-crafted features of  time-series photometry and metadata with decision forests in a supervised context \citep{2007A&A...475.1159D,2011ApJ...733...10R,2011MNRAS.414.2602D,2013AJ....146..101P}. Neural network approaches to learn representations of time-series photometry (both in supervised and self-supervised contexts) then achieved state of the art, first with flavors of RNNs (e.g., LSTMs: \citealt{2018NatAs...2..151N}, GRUs: \citealt{2019PASP..131k8002M,2020MNRAS.493.2981B}) and more recently with convolution \citep{2020ApJS..250...30J,2021AJ....162..275B} and Transformers \citep{2023A&A...670A..54D,2024MNRAS.527.1494L}. CNNs have been used to achieve state of the art classification on galaxy spectra (e.g., GalSpecNet: \citealt{2024MNRAS.527.1163W}). \citet{2021ApJ...911L..33H} use CNN autoencoders with contrastive learning for self-supervised embedding of galaxy images.

AstroCLIP \citep{Parker_2024} fused pre-trained embeddings of galaxy spectra and images with constrastive learning and showed the trained model to be competitive with purpose-built classification models. Our work differs from AstroCLIP in that 1) our primary objects are individual sources that vary in time (i.e. not static like galaxies); 2) we explicitly build embeddings for three different modes of data; 3) our approach does not rely upon pretraining of embeddings for the different modes, but instead learns all embeddings simultaneously; and 4) we examine the efficacy of the model with missing modes at test time. Like with AstroCLIP, we find our model outperforms purpose-built supervised models for downstream tasks. To our knowledge, MAVEN \citep{2024arXiv240816829Z} is the only other CLIP-centric model applied in the astronomical time domain. It is a dual-mode model built for ``one off'' explosive supernovae events, whereas ours is focused on persistently variable sources. MAVEN first learns spectroscopic and photometric embeddings from synthetic data and then requires a fine-tuning step on real survey data. Our model is trained directly on real observational data. 

\section{Dataset Assembly}
\label{sec:dataset}

The basis of our observational dataset is the variable star catalog \citep{jayasinghe2024var} observed and curated by the All-Sky Automated Survey for SuperNovae (ASAS-SN) project \citep{shappee2014asassn}. We downloaded the lightcurve data from the 2021 assembly of the 687,695 {\it v}-band variables and the 2022 assembly of the 378,861 {\it g}-band variables, along with the associated metadata catalogs. These catalogs contain cross-matched photometry information for each source from WISE \citep{2010AJ....140.1868W}, GALEX \citep{2007ApJS..173..682M}, 2MASS \citep{2006AJ....131.1163S} and Gaia EDR3 \citep{2021A&A...649A...1G}, variability statistics derived from the lightcurves in each bandpass (such as period and peak-to-peak amplitude), astrometric information from Gaia (such as parallax and proper motion), and a machine-learned classification from the ASAS-SN group \citep{jayasinghe2024var}. We deduplicated and merged these data using the cross-matched \texttt{source\_id} from Gaia EDR3, with the merged catalog serving as the basis of the \texttt{metadata} mode. 

To facilitate the use of positional information in the models, we transformed the galactic latitude $b\rightarrow \sin(b)$ and galactic longitude $l\rightarrow \cos(l)$. We also transformed all catalog apparent photometry $m$ to absolute magnitude using the Gaia EDR3 parallax $\pi$ (units of milliarcseconds) using $M = m + 5 \log_{10} \pi - 10$. We did not deredderen any values. To cleanly delineate the \texttt{time-series} mode from the \texttt{metadata} mode, we removed features derived from photometric time-series data from the \texttt{metadata} catalog (and later used such features as auxiliary inputs in the \texttt{time-series} channel, see \ref{sec:phot} below). We also removed any columns from the \texttt{metadata} catalog related to indices (such as source names).  Last, we removed the assigned classification of each source (later used to test downstream tasks; see \ref{sec:results}).

To build the \texttt{spectral} mode, we cross-matched the sources with the v2.0 DR9 Large Sky Area Multi-Object Fiber Spectroscopic Telescope (LAMOST; \citealt{2012RAA....12.1197C}) public catalog using the Gaia EDR3 ID. We downloaded the 41,204 1D spectra identified in the the cross match and constructed a lookup table matching specific variable sources to LAMOST spectra. Most variable sources had zero associated spectra but a small subset had multiple spectra of the same source obtained over multiple epochs.

We filtered the dataset based on the following criteria: (1) each object must have data available for all three modalities—time-series photometry, spectra, and metadata; (2) the metadata cannot have any missing values to ensure a complete dataset for training; and (3) the object must belong to one of the top 10 classes to ensure there are sufficient samples for effective CLIP training \citep{xu2023demystifying, alabdulmohsin2024clip}. The selected classes and the corresponding number of objects are listed in Table \ref{table:dataset}.

\begin{table}
\centering
\small  % Use a smaller font size
\begin{tabular}{@{}l l c@{}}
\toprule
\textbf{Class} & \textbf{Description} & \textbf{Total Objects} \\
\midrule
\textbf{EW}    & W Ursae Majoris type binaries & 6168 \\
\textbf{SR}    & Semi-regular variables & 4590 \\
\textbf{EA}    & Detached Algol-type binaries & 2916 \\
\textbf{RRAB}  & Fundamental Mode RR Lyrae variables & 2351 \\
\textbf{EB}    & $\beta$ Lyrae-type binaries & 1976 \\
\textbf{ROT}   & Spotted Variables with rotational modulation & 1839 \\
\textbf{RRC}   & First Overtone RR Lyrae variables & 796 \\
\textbf{HADS}  & High amplitude $\delta$ Scuti type variables & 281 \\
\textbf{M}     & Mira variables & 268 \\
\textbf{DSCT}  & $\delta$ Scuti type variables & 255 \\
\bottomrule
\end{tabular}
\caption{Summary of variable star classes, including abbreviations, descriptions, and total object counts for each class used in the dataset.}
\label{table:dataset}
\end{table}

\section{Method}

Our objective is to develop a self-supervised multimodal model that can learn from astronomical data across three distinct modalities: time-series photometry, spectra, and astrophysical metadata. To achieve this, we extend the Contrastive Language-Image Pretraining (CLIP) framework \citep{radford2021learning} to a trimodal setting, enabling simultaneous learning from multiple data types. In this section, we describe the models used for each modality and how they are integrated into our multimodal CLIP framework.

\subsection{Photometric Time-Series Model}
\label{sec:phot}

Photometric time-series data are flux measurements of astronomical objects over time. To effectively capture the temporal dependencies and handle sequences of varying lengths, we employ the Encoder component from the Informer model \citep{zhou2021informer}.

\vspace{1em}
\textbf{Model Architecture.} The photometric time-series encoder consists of:

\begin{itemize} 
\item {Input Embedding Layer:} Projects the input features to a higher-dimensional space. 
\item {Informer Encoder Layers:} Eight encoder layers with a hidden dimension of 128, four attention heads, and a feedforward dimension of 512.
\item {Output Layer:} Produces a fixed-length embedding representing the input time-series data. \end{itemize}

\textbf{Data Preprocessing.} Each light curve is a sequence of flux measurements \( f = \{f_1, f_2, \dots, f_T\} \) and flux errors \( \sigma_f = \{\sigma_{f_1}, \sigma_{f_2}, \dots, \sigma_{f_T}\} \) at corresponding times \( t = \{t_1, t_2, \dots, t_T\} \). We normalize the flux by subtracting the mean \( \mu_f \) and dividing by the median absolute deviation \( \text{MAD}_f \): \( \tilde{f}_i = \frac{f_i - \mu_f}{\text{MAD}_f} \). Flux errors are normalized by the flux median absolute deviation division: \( \tilde{\sigma}_{f_i} = \frac{\sigma_{f_i}}{\text{MAD}_f} \). Time is scaled between 0 and 1 for each light curve: 
\(\delta_{t} = t_{\max} - t_{\min}\); \(\tilde{t}_i = \frac{t_i - t_{\min}}{\delta_{t}}\). Auxiliary features such as amplitude, period, Lafler-Kinmann string length statistic \citep{1965ApJS...11..216L}, peak-to-peak variability, delta time $\frac{\delta_{t}}{365}$ and logarithm of median absolute deviation \( \log \text{MAD}_f \) are included as additional inputs.

\vspace{1em}
\textbf{Handling Variable Sequence Lengths.} We set a maximum sequence length of $L = 200$. Sequences longer than this are randomly cropped during training and center-cropped during validation and testing. Shorter sequences are padded with zeros, and an attention mask is used to differentiate between valid data and padding.

\subsection{Spectra Model}

Spectral data provides detailed information about the composition and physical properties of astronomical objects. We adapt the \hbox{GalSpecNet} architecture \citep{wu2024galaxy}, which is specifically designed for processing one-dimensional astronomical spectra.

\vspace{1em}
\textbf{Model Architecture.} The spectra encoder consists of:

\begin{itemize} 
\item {Convolutional Layers:} Four layers (64, 64, 32, 32 channels) followed by ReLU activations. 
\item {Pooling Layers:} Max-pooling layers after each convolutional layer except for the last one. 
\item {Dropout Layer:} Applied after the last convolutional layer for regularization. 
\item {Output Layer:} Generates a fixed-length embedding of the spectral data. 
\end{itemize}

\vspace{1em}
\textbf{Modifications.} We reduce the last three fully connected layers to a single one for classification or omit it entirely when using the model as a feature extractor. We also add additional input channels for spectra errors and auxiliary data.

\vspace{1em}
\textbf{Data Preprocessing.} Spectra are limited to the wavelength range of 3850–9000 Å and resampled at regular intervals of 2Å using linear interpolation. Each spectrum \( s = \{s_1, s_2, \dots, s_W\} \) and its uncertainties \( \sigma_s = \{\sigma_{s_1}, \sigma_{s_2}, \dots, \sigma_{s_W}\} \) at corresponding wavelengths \( w = \{w_1, w_2, \dots, w_W\} \) are normalized in a similar way as photometry data: values are normalized by subtracting the mean \( \mu_s \) and dividing by the median absolute deviation \( \text{MAD}_s \): \( \tilde{s}_i = \frac{s_i - \mu_s}{\text{MAD}_s} \), while uncertainties are divided by \( \text{MAD}_s \): \( \tilde{\sigma}_{s_i} = \frac{\sigma_{s_i}}{\text{MAD}_s} \). The logarithm of the median absolute deviation \( \log \text{MAD}_s \) is included as an auxiliary feature.

\subsection{Metadata Model}

The metadata modality consists of astrophysical parameters and observational data not included in the other two modalities. This includes features like absolute magnitudes in various bands, astrometric information, and other cross-matched catalog data. A full list of features and their descriptions is provided in Table     \ref{table:feature_descriptions}.

\vspace{1em}
\textbf{Model Architecture.} The metadata encoder % (Figure \ref{fig:meta}) 
is a Multilayer Perceptron consisting of:

\begin{itemize} 
\item {Input Layer:} Accepts the 34 preprocessed features. 
\item {Hidden Layers:} Two hidden layers with 512 units each followed by ReLU activations. 
\item {Dropout Layers:} Applied after hidden layers for regularization. 
\item {Output Layer:} Provides a fixed-length metadata embedding. \end{itemize}

\vspace{1em}
\textbf{Data Preprocessing.} Except for the steps already mentioned during the dataset assembly (see \ref{sec:dataset}), we apply logarithm to period and then standardize each feature to have zero mean and unit variance.

\subsection{AstroM\texorpdfstring{$^3$}{3}: Multi-modal CLIP Model}

To integrate the three modalities we extend the CLIP model to a trimodal setting and name the entire architectural approach as {\bf AstroM}$\mathbf{^3}$. The goal is to learn a shared embedding space where representations from different modalities corresponding to the same astronomical object are close together (see Figure~\ref{fig:clip}).

\vspace{1em}
\textbf{Projection Heads.} Each modality has its own architecture, producing embeddings of different sizes. To bring these embeddings into a shared space, we apply a projection head to each modality. The projection head is a fully connected layer that maps the embeddings to a fixed size of 512. Let the original embeddings of photometry, spectra, and metadata be denoted as $\tilde{P}_i$, $\tilde{S}_i$, and $\tilde{M}_i$, where $i$ denotes the $i$-th sample in a batch of size $N$. The projection heads transform these original embeddings as follows:
\begin{align}
    P_i &= W_P \tilde{P}_i + b_P \\
    S_i &= W_S \tilde{S}_i + b_S \\
    M_i &= W_M \tilde{M}_i + b_M,
\end{align}

\noindent where $W_P$, $W_S$, and $W_M$ are the weight matrices, and $b_P$, $b_S$, and $b_M$ are the bias terms for the projection head of each modality. After applying these transformations, the projected embeddings $P_i$, $S_i$, and $M_i$ all have a fixed size of $512$, making them suitable for comparison in the shared embedding space.

\vspace{1em}
\textbf{Pairwise Similarity Matrices.} 
For each pair of modalities (photometry-spectra, spectra-metadata, metadata-photometry) we compute similarity matrices using cosine similarity:
\begin{align}
{PS}_{ij} &= \frac{P_i \cdot S_j}{\|P_i\| \|S_j\|} \\
{SM}_{ij} &= \frac{S_i \cdot M_j}{\|S_i\| \|M_j\|} \\
{MP}_{ij} &= \frac{M_i \cdot P_j}{\|M_i\| \|P_j\|}
\end{align}

\textbf{Contrastive Loss.} 
We use a symmetric cross-entropy loss to align the embeddings:
\begin{align}
\mathcal{L}^{PS} &= \mathcal{L}_{\text{CE}}({PS}, {Y}) + \mathcal{L}_{\text{CE}}({PS^\top}, {Y}) \\
\mathcal{L}^{SM} &= \mathcal{L}_{\text{CE}}({SM}, {Y}) + \mathcal{L}_{\text{CE}}({SM^\top}, {Y}) \\
\mathcal{L}^{MP} &= \mathcal{L}_{\text{CE}}({MP}, {Y}) + \mathcal{L}_{\text{CE}}({MP^\top}, {Y})
\end{align}

where $\mathcal{L}_{\text{CE}}$ denotes the cross-entropy loss and ${Y}$ is the label matrix defined as:
\begin{align}
Y_{ij} = 
\begin{cases} 
1 & \text{if } i = j, \\
0 & \text{otherwise}.
\end{cases}
\end{align}

\textbf{Total Loss.} The overall loss is the sum of the individual pairwise losses:
\begin{align}
\mathcal{L} = \mathcal{L}^{PS} + \mathcal{L}^{SM} + \mathcal{L}^{MP}
\end{align}

By minimizing this loss, the model learns to align the embeddings across all three modalities, bringing representations of the same object closer together in the embedding space while pushing apart those of different objects.

\section{Results}
\label{sec:results}

\begin{table}
\centering
\begin{tabular}{@{}l S[table-format=2.3]@{\,\( \pm \)\,} S[table-format=1.3] S[table-format=2.3]@{\,\( \pm \)\,} S[table-format=1.3]@{}}
\toprule
\textbf{Data Type} & \multicolumn{2}{c}{\textbf{No CLIP}} & \multicolumn{2}{c}{\textbf{CLIP}} \\ 
\midrule
\textbf{Spectra}   & 76.278 & 0.931 & 77.396 & 0.614 \\
\textbf{Metadata}      & 85.623 & 0.628 & 85.855 & 0.856 \\
\textbf{Photometry}     & 84.642 & 6.317 & \textbf{91.468} & \textbf{0.446} \\
\textbf{All}       & 94.065 & 0.390 & 94.153 & 0.577 \\
\bottomrule
\end{tabular}
\caption{Classification accuracy comparison between models trained with and without CLIP. Statistically significant improvements are in bold.}
\label{table:clip_comparison}
\end{table}

We evaluated the models on downstream classification across four modes: photometry only, spectra only, metadata only, and all modalities combined. For single modalities, we added a fully connected layer on top of the respective encoders for classification. In the multimodal setting, we averaged the embeddings from all three modalities and then applied a fully connected layer for classification. Each model was trained in two ways: with CLIP pre-training, where the model was initially trained using the CLIP framework and then fine-tuned for the downstream task, and without CLIP pre-training, where models were trained directly on the task with randomly initialized weights. Importantly, model architecture and setup were identical across all conditions, differing only in the initialization of weights. The training setup and hyperparameter search process are detailed in Appendix \ref{sec:hyperparameters}. All models were cross-validated using 5 random seeds and data splits for robust evaluation.

\begin{table*}
\centering
\small  % Use a smaller font size
\begin{tabular}{@{}l cccc cccc cccc@{}}
\toprule
\multirow{2}{*}{\textbf{Class}} & \multicolumn{4}{c}{\textbf{Train}} & \multicolumn{4}{c}{\textbf{Val}} & \multicolumn{4}{c}{\textbf{Test}} \\
\cmidrule(lr){2-5} \cmidrule(lr){6-9} \cmidrule(lr){10-13}
 & \textbf{Full} & \textbf{50\%} & \textbf{25\%} & \textbf{10\%} & \textbf{Full} & \textbf{50\%} & \textbf{25\%} & \textbf{10\%} & \textbf{Full} & \textbf{50\%} & \textbf{25\%} & \textbf{10\%} \\
\midrule
\textbf{EW}    & 4890 & 1209 & 516 & 166 & 597  & 149  & 64   & 21   & 681  & 160  & 69   & 22  \\
\textbf{SR}    & 3647 & 1209 & 516 & 166 & 479  & 149  & 64   & 21   & 464  & 160  & 69   & 22  \\
\textbf{EA}    & 2343 & 1209 & 516 & 166 & 272  & 149  & 64   & 21   & 301  & 160  & 69   & 22  \\
\textbf{RRAB}  & 1886 & 1209 & 516 & 166 & 231  & 149  & 64   & 21   & 234  & 160  & 69   & 22  \\
\textbf{EB}    & 1571 & 1209 & 516 & 166 & 207  & 149  & 64   & 21   & 198  & 160  & 69   & 22  \\
\textbf{ROT}   & 1454 & 1209 & 516 & 166 & 189  & 149  & 64   & 21   & 196  & 160  & 69   & 22  \\
\textbf{RRC}   & 624  & 624  & 516 & 166 & 93   & 93   & 64   & 21   & 79   & 79   & 69   & 22  \\
\textbf{HADS}  & 226  & 226  & 226 & 166 & 29   & 29   & 29   & 21   & 26   & 26   & 26   & 22  \\
\textbf{M}     & 216  & 216  & 216 & 166 & 30   & 30   & 30   & 21   & 22   & 22   & 22   & 22  \\
\textbf{DSCT}  & 206  & 206  & 206 & 166 & 25   & 25   & 25   & 21   & 24   & 24   & 24   & 22  \\
\bottomrule
\end{tabular}
\caption{Class distribution across training, validation, and test sets for different dataset splits (Full, 50\%, 25\%, 10\%), created by downsampling the most common classes to balance subsets.}
\label{table:class_distribution}
\end{table*}

\begin{table*}
\centering
\begin{tabular}{@{}l l S[table-format=2.3]@{\,\( \pm \)\,} S[table-format=1.3] S[table-format=2.3]@{\,\( \pm \)\,} S[table-format=1.3] S[table-format=2.3]@{\,\( \pm \)\,} S[table-format=1.3]@{}}
\toprule
\textbf{Data Type} & \textbf{Pre-train} & \multicolumn{2}{c}{\textbf{50\%}} & \multicolumn{2}{c}{\textbf{25\%}} & \multicolumn{2}{c}{\textbf{10\%}} \\
\midrule
\textbf{Spectra}   & No CLIP & 68.072  & 1.759  & 63.729  & 1.637  & 46.677  & 3.486 \\
                   & CLIP    & 71.609  & 1.814  & \textbf{67.869}  & \textbf{1.303}  & \textbf{59.235}  & \textbf{1.399} \\
\addlinespace
\textbf{Photometry} & No CLIP & 89.177  & 0.518  & 83.218  & 2.709  & 83.073  & 1.762 \\
                   & CLIP    & 90.272  & 0.695  & \textbf{88.434}  & \textbf{0.781}  & \textbf{90.720}  & \textbf{1.359} \\
\addlinespace
\textbf{Metadata}   & No CLIP & 82.035  & 1.452  & 79.649  & 1.148  & 76.524  & 1.309 \\
                   & CLIP    & 83.830  & 1.083  & 81.953  & 1.492  & 79.073  & 1.711 \\
\addlinespace
\textbf{All}        & No CLIP & 91.870  & 0.470  & 90.741  & 1.053  & 88.264  & 2.188 \\
                   & CLIP    & 91.978  & 0.746  & 92.073  & 1.066  & 90.628  & 1.509 \\
\bottomrule
\end{tabular}
\caption{Accuracy comparison across data splits (50\%, 25\%, 10\%) with and without CLIP pre-training for different data types (Spectra, Photometry, Metadata, All). Statistically significant improvements in bold.}
\label{table:split_comparison}
\end{table*}

\subsection{CLIP Evaluation}

The results in Table \ref{table:clip_comparison} show that while there is no statistically significant difference between using CLIP and not using CLIP for spectra, metadata and combined modalities, CLIP has a strong impact on photometry classification. It increased the average accuracy \textbf{from 84.64\% to 91.47\%} and  significantly reduced the standard deviation (from 6.32 to 0.45), indicating better model stability. With or without CLIP, we also show that \textit{by using all three modalities at the same time, we achieve better accuracy than by using any single modality alone}.

\subsection{Limited Labeled Data}

To evaluate the effectiveness of CLIP pre-training when the availability of labeled data is limited, we conducted experiments on smaller subsets of the original dataset. Specifically, we created subsets containing 10\%, 25\%, and 50\% of the data by downsampling the most common classes, ensuring a balanced class distribution. Table \ref{table:class_distribution} provides details on the class distribution across these subsets. Note that we choose to downsample the overrepresented sources at random. An interesting alternative to this, to approximate the ways in which brighter sources preferentially are easier to label on new survey data, would be to select only the brightest (or highest signal-to-noise) sources to include in the training data.

\begin{figure*}
    \centering
    \begin{subfigure}{0.49\textwidth}
        \centering
        \ifdownsampled
            \includegraphics[width=\textwidth]{images/umap-train-down.png} % Low-resolution version
        \else
            \includegraphics[width=\textwidth]{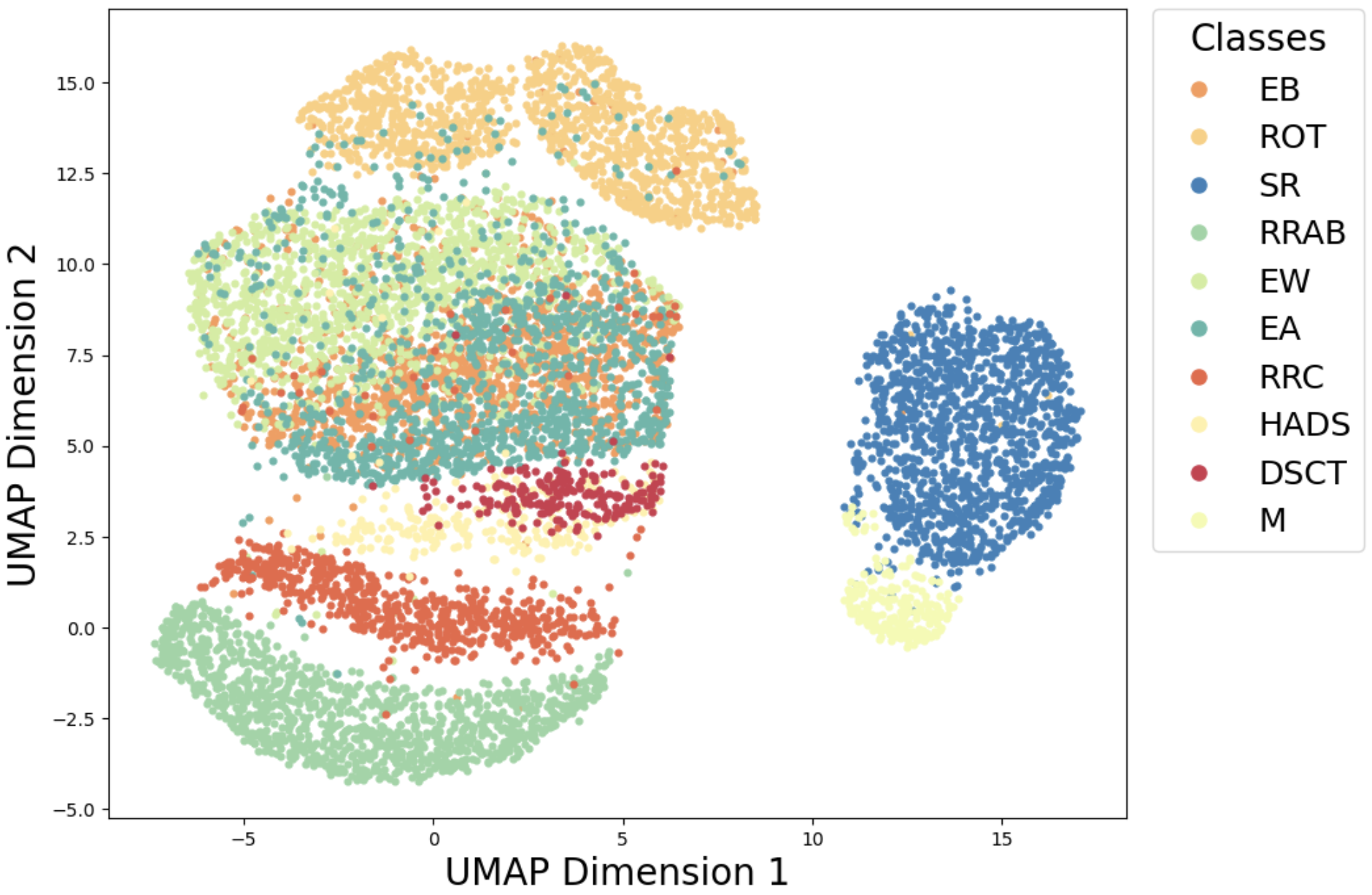} % Full-resolution version
        \fi
        \caption{UMAP - Train}
        \label{fig:umap-train}
    \end{subfigure}
    \hfill
    \begin{subfigure}{0.49\textwidth}
        \centering
        \ifdownsampled
            \includegraphics[width=\textwidth]{images/umap-test-down.png} % Low-resolution version
        \else
            \includegraphics[width=\textwidth]{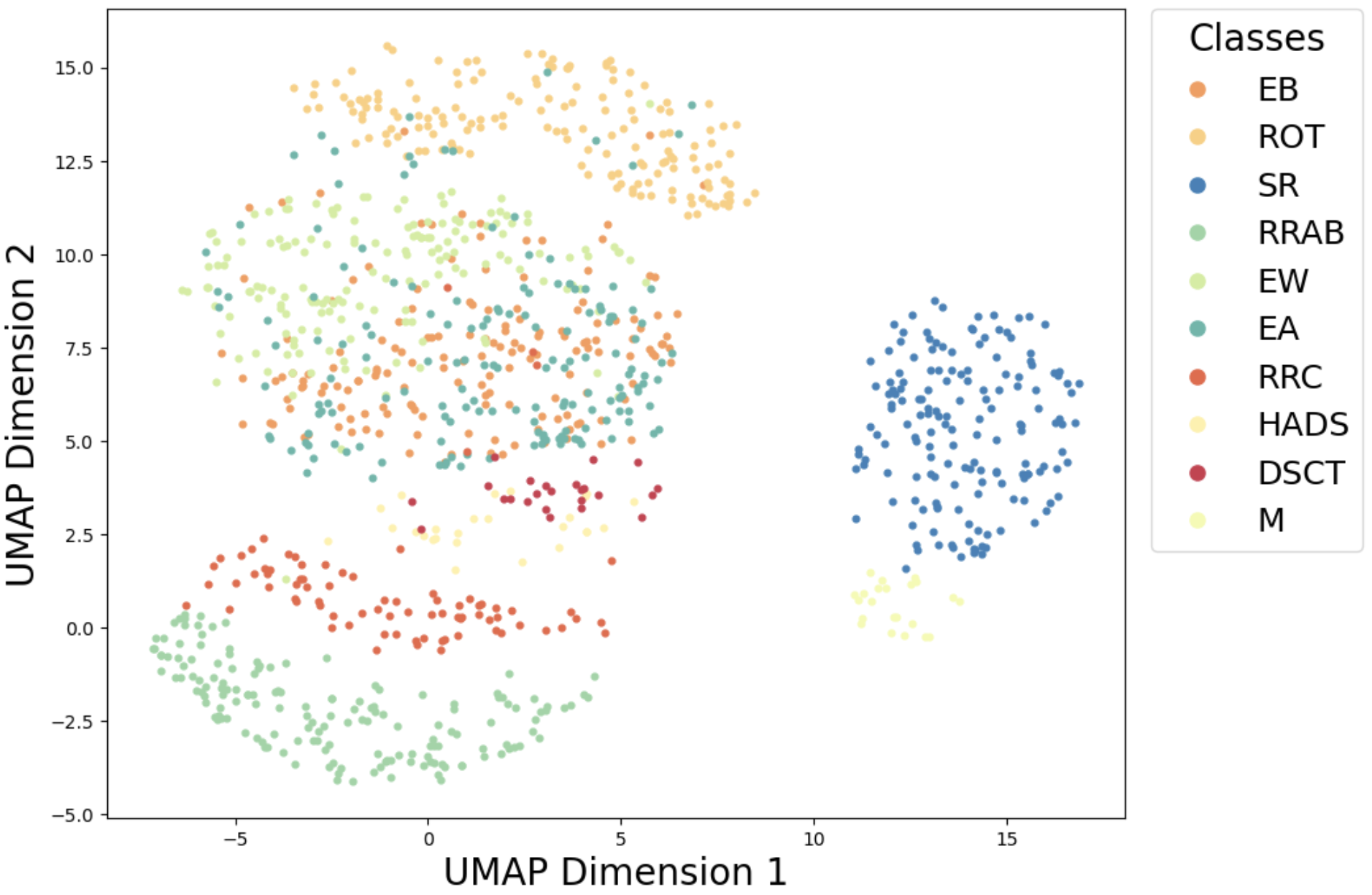} % Full-resolution version
        \fi
        \caption{UMAP - Test}
        \label{fig:umap-test}
    \end{subfigure}
    \caption{UMAP visualizations of multimodal embeddings: (a) training set and (b) test set, showing class separability and alignment between sets. Each source in the training and test set are coloured by the class determined in \citep{jayasinghe2024var} but these class labels are not used in the construction of the embeddings.}
    \label{fig:umap}
\end{figure*}

\vspace{1em}
\textbf{Models.} For each subset, we retrained all models, with and without CLIP pre-training, using the same optimization settings and hyperparameter search as previously applied. It is important to note that the CLIP model used for these experiments was the same as before: pre-trained on the full dataset without using any labels. This setup is designed (for future applications) to leverage large amounts of unlabeled data for pre-training and then fine-tuning the model on smaller labeled datasets.

\vspace{1em}
\textbf{Results.} The results in Table \ref{table:split_comparison} demonstrate that CLIP pre-training improves model performance when labeled data is limited. For example, at the 25\% data split, CLIP increased the accuracy of the spectra model by \textbf{4.14\%} (from 63.73\% to 67.87\%), and by \textbf{12.56\%} at the 10\% data split (from 46.68\% to 59.24\%). Photometry shows a similar trend, with accuracy increasing by \textbf{5.21\%} at the 25\% data split (from 83.22\% to 88.43\%), and by \textbf{7.65\%} at the 10\% split (from 83.07\% to 90.72\%). For metadata and all modalities combined, although the difference in accuracy between models with and without CLIP pre-training was not statistically significant, CLIP models generally performed better. These findings suggest that CLIP is beneficial, especially when labeled training data is limited, making it an effective approach for leveraging large unlabeled datasets in future work.

\begin{figure*}
    \ifsvg
        \includesvg[width=\textwidth]{images/outliers}
    \else
        \includegraphics[width=\textwidth]{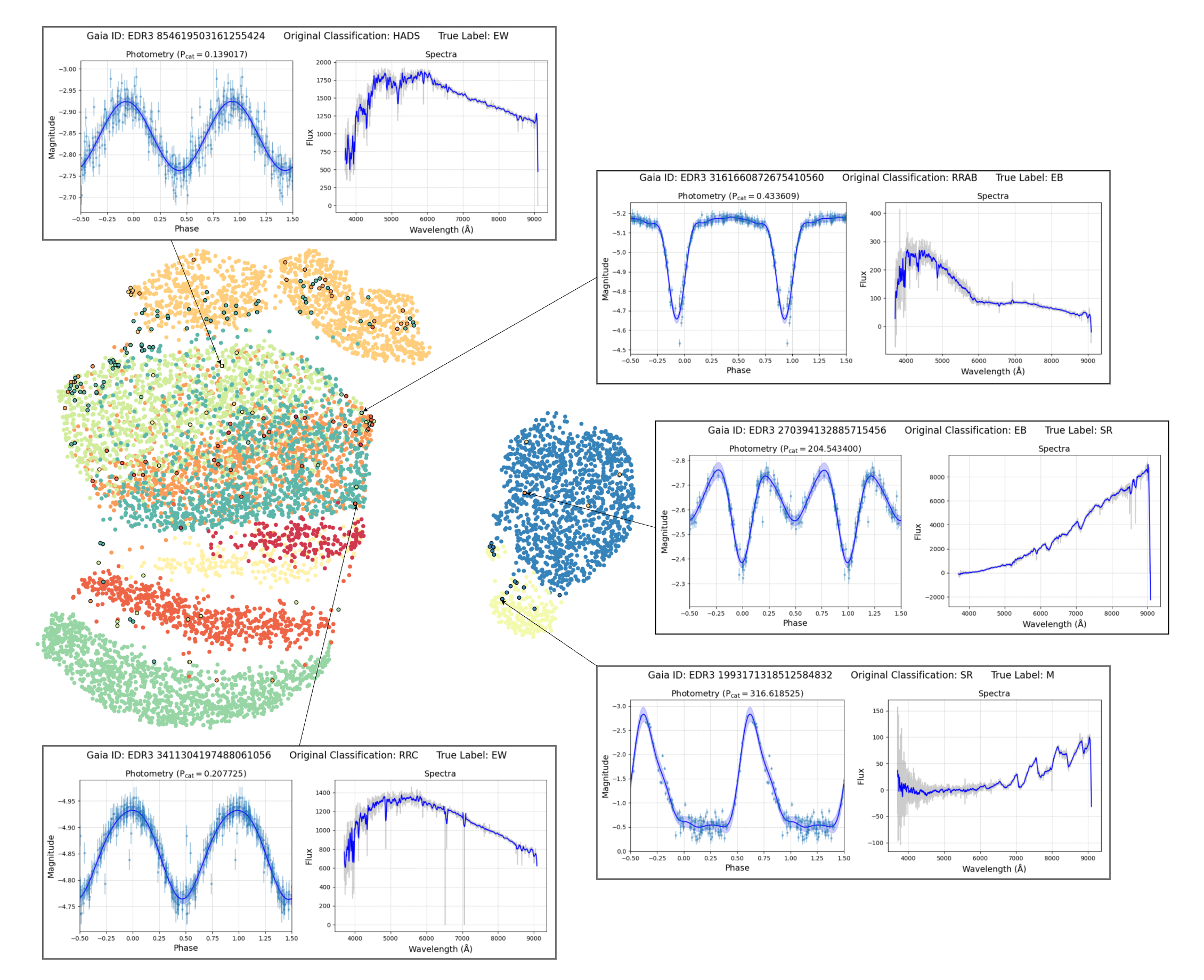}
    \fi
    \caption{Examples of catalog misclassifications with photometry and spectrum for each object. Top to bottom: (1) Likely EW missclassified as HADS; (2) V* AC CMi, a known semi-detached binary misclassified as RR Lyrae; (3) Possible SR or Mira variable with period alignment issues; (4) Known Mira variable (V0439 Cas) misclassified as SR; (5) Likely EW binary \citep{2023A&A...674A..16M} misclassified as RRC.}
    \label{fig:outliers}
\end{figure*}

\begin{figure*}
    \centering
    \begin{subfigure}{0.49\textwidth}
        \centering
        \ifdownsampled
            \includegraphics[width=\textwidth]{images/spec898-down.png} % Low-resolution version
        \else
            \includegraphics[width=\textwidth]{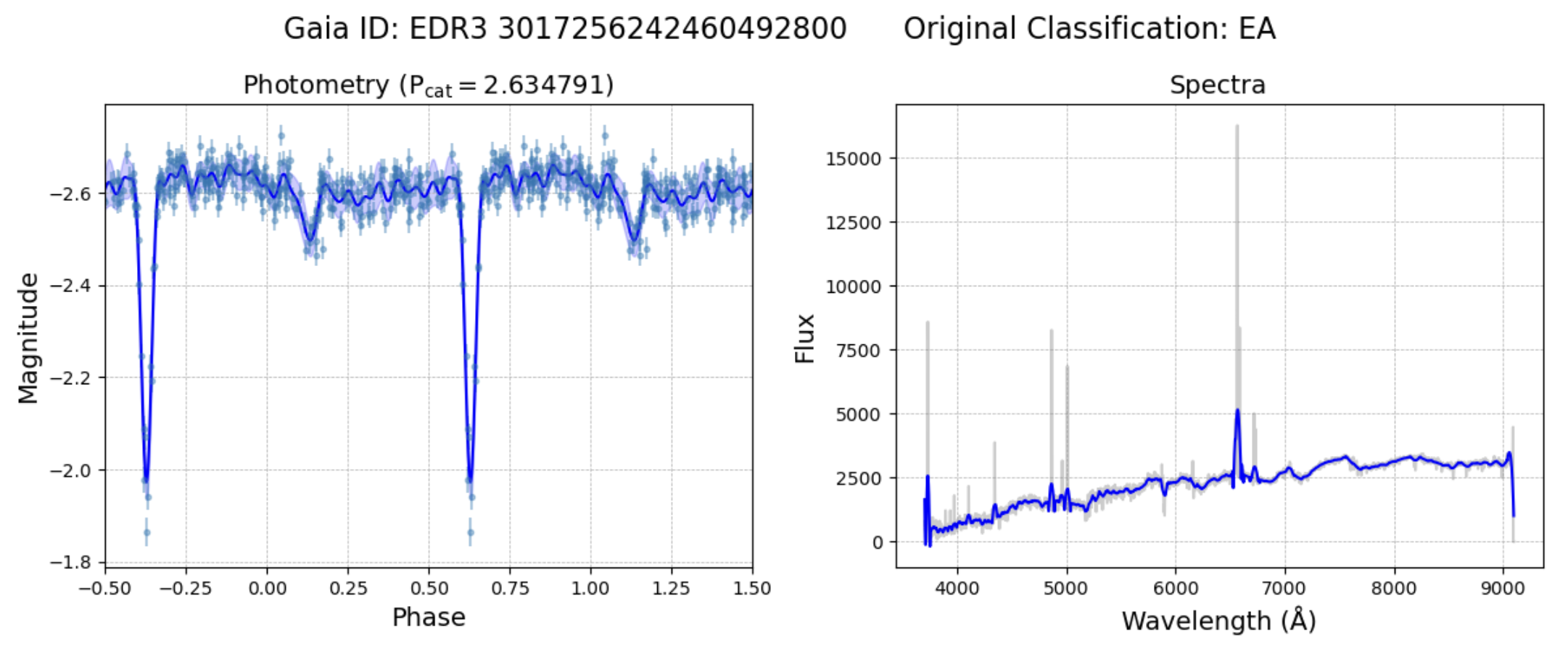} % Full-resolution version
        \fi
        \caption{}
        \label{fig:inclass1}
    \end{subfigure}
    \hfill
    \begin{subfigure}{0.49\textwidth}
        \centering
        \ifdownsampled
            \includegraphics[width=\textwidth]{images/spec1159-down.png} % Low-resolution version
        \else
            \includegraphics[width=\textwidth]{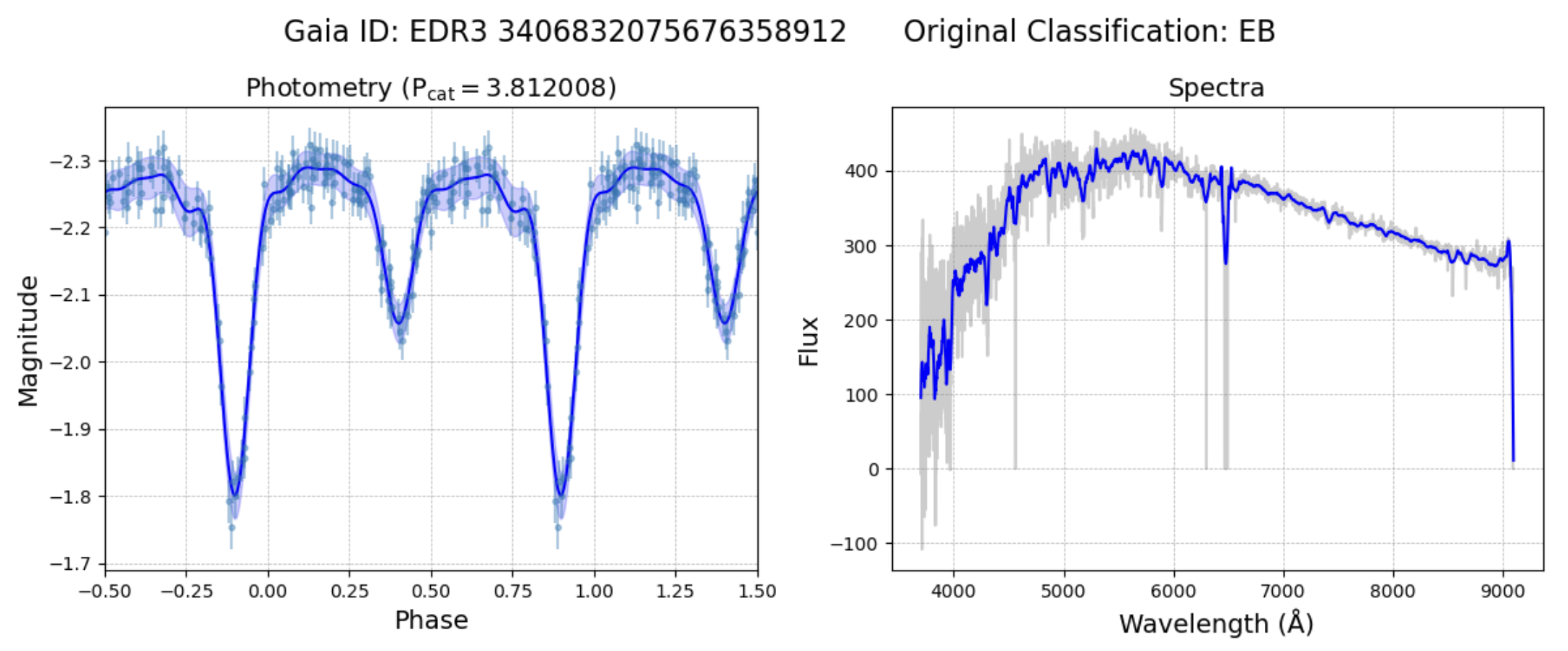} % Full-resolution version
        \fi
        \caption{}
        \label{fig:inclass2}
    \end{subfigure}
    \vspace{0.5em} % Adds vertical spacing between rows
    \begin{subfigure}{0.49\textwidth}
        \centering
        \ifdownsampled
            \includegraphics[width=\textwidth]{images/spec1235-down.png} % Low-resolution version
        \else
            \includegraphics[width=\textwidth]{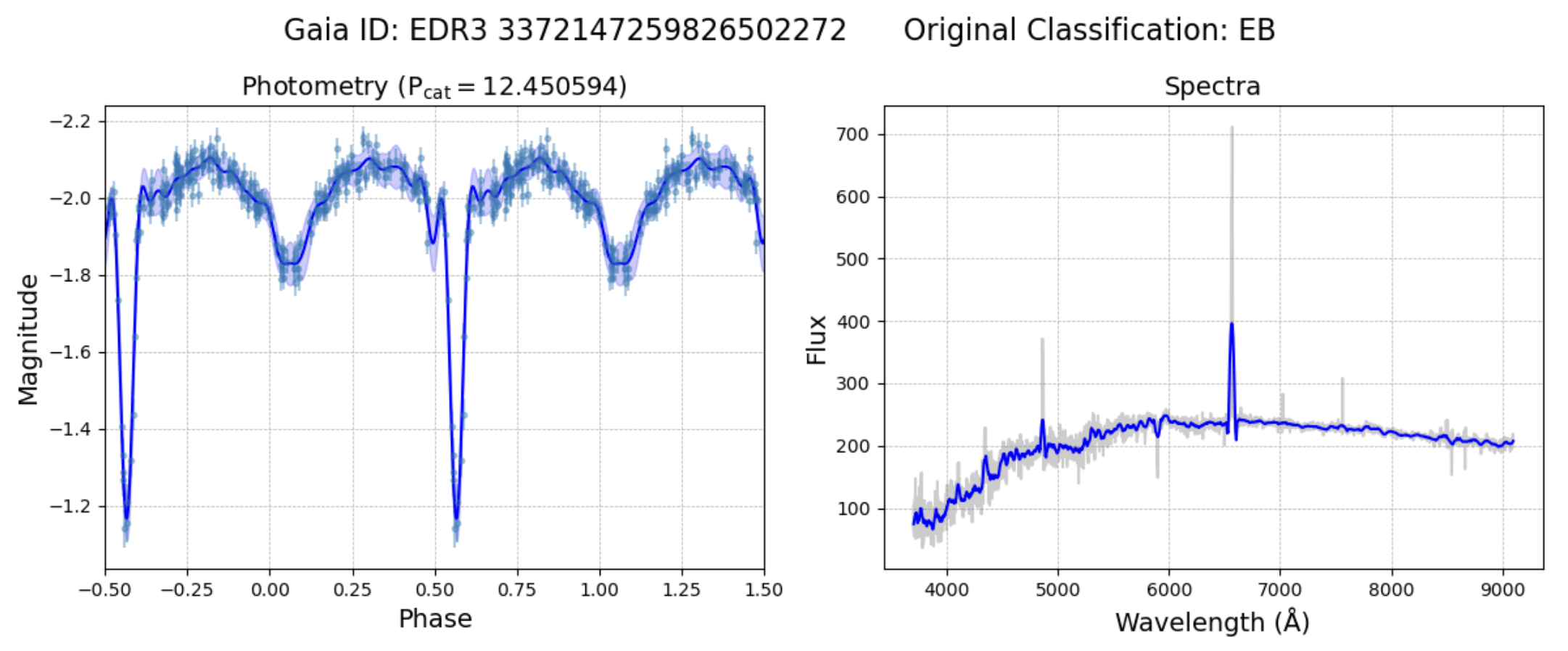} % Full-resolution version
        \fi
        \caption{}
        \label{fig:inclass3}
    \end{subfigure}
    \hfill
    \begin{subfigure}{0.49\textwidth}
        \centering
        \ifdownsampled
            \includegraphics[width=\textwidth]{images/spec6265-down.png} % Low-resolution version
        \else
            \includegraphics[width=\textwidth]{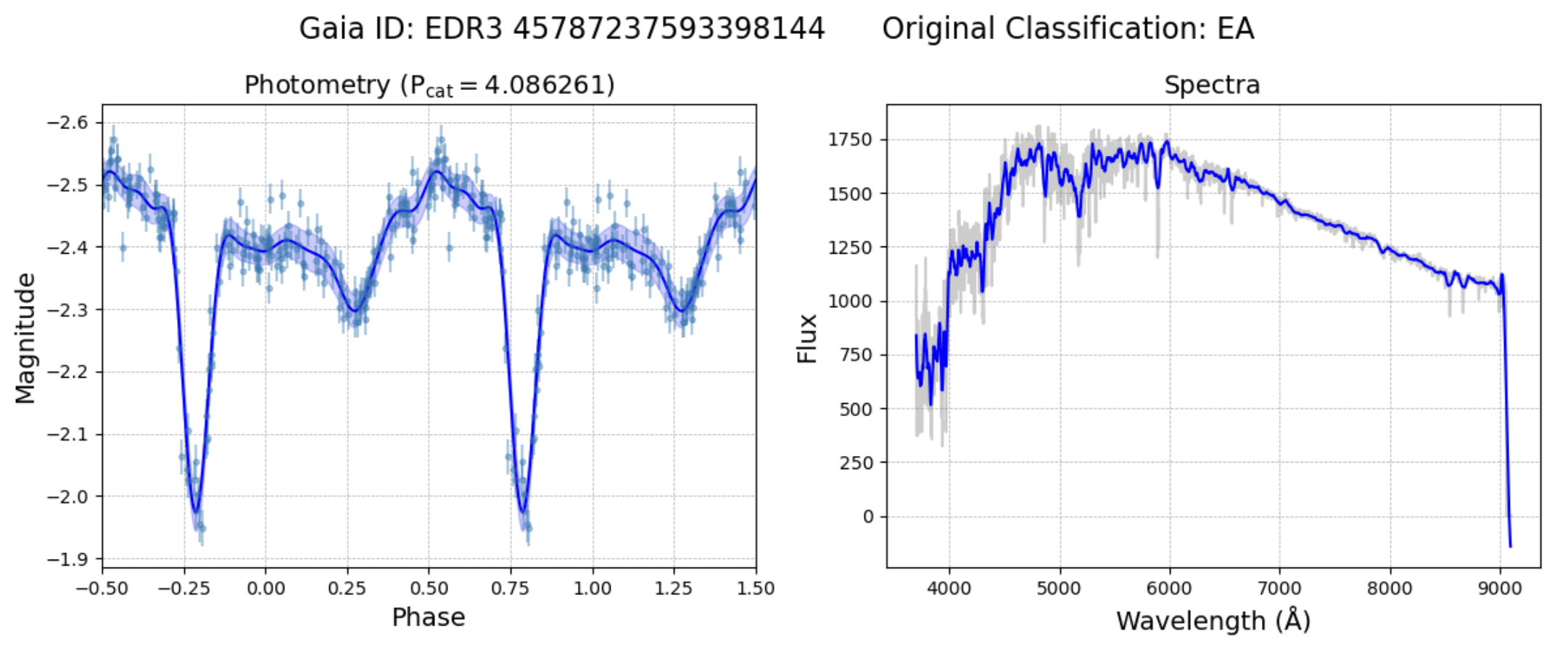} % Full-resolution version
        \fi
        \caption{}
        \label{fig:inclass4}
    \end{subfigure}
    \caption{Examples of in-class outliers flagged by the model due to distinctive features, despite correct labels. (a) EA-type star, V1174 Ori, an X-ray bright pre-main sequence system \citep{2022ApJ...941..125S}. (b) EB-type star with unusual out-of-eclipse modulations, possibly due to rotation. (c) Semi-detached binary with emission lines. (d) Likely an EB misclassified as EA, with light curve patterns indicating rotation or pulsation.}
    \label{fig:inclass}
\end{figure*}

\subsection{UMAP Analysis}

We use Uniform Manifold Approximation and Projection (UMAP) method \citep{mcinnes2018umap} to visualize how well our model distinguishes among classes in the embedding space. UMAP is fit on the averaged embeddings across all modalities from the training set, and projections are generated for both the training (Figure \ref{fig:umap-train}) and the test (Figure \ref{fig:umap-test}) sets. The results show that:
\begin{itemize}
\item Most classes are well separated, though Detached Algol-type binaries (EA), $\beta$ Lyrae-type binaries (EB) and W Ursae Majoris type binaries (EW) partially overlap. This is expected on a physical basis, as these are all types of binary stars and share similar characteristics\footnote{Our photometry embedding works in time-flux space; future extensions to phase-flux space may help disambiguate the eclipsing binary classes.}.
\item As expected, the test set follows the same UMAP projection structure as the training set. For instance, Spotted Variables with rotational modulation (ROT) from the test set align with their counterparts in the training set.
\end{itemize}

\vspace{1em}
\textbf{Outliers}. Based on the UMAP projections, we observed that some objects were located outside their expected clusters. To investigate further, we trained a DBSCAN model \citep{ester1996density} on each class, configuring it to identify a single major cluster per class, with all objects outside of that cluster marked as outliers. We manually reviewed the objects flagged as outliers and found that most objects are falling into two categories: (1) objects with incorrectly assigned classifications from the catalog, and (2) objects with the correct labels that are in-class outliers  because of their unique features.

\vspace{1em}
\textbf{Misclassifications}. Figure~\ref{fig:outliers} highlights misclassification candidates, showing both the photometry and spectrum for representative examples summarized below:
\begin{itemize}
    \item EDR3 854619503161255424, Likely EW Binary: The reported Gaia period is twice that of the catalog (0.2780335 days), suggesting this source is likely an EW binary. The lack of the asymmetric shape typical of a High Amplitude Delta Scuti (HADS) star supports this reclassification.

    \item EDR3 3161660872675410560, Semi-detached Binary (EB): This source, V* AC CMi, is a known semi-detached binary (EB), suggesting that the RR Lyrae (RRL) classification is incorrect.

    \item EDR3 270394132885715456, Possible SR or Mira Variable: Gaia lists half the period (102 days) compared to the catalog, but the catalog period appears correct. An SR or Mira classification is likely more appropriate.

    \item EDR3 1993171318512584832, Known Mira Variable: This source, V0439 Cas, is a known Mira variable, indicating that its current SR classification is inaccurate.

    \item EDR3 3411304197488061056, Likely EW Binary with Incorrect Catalog Period: Gaia classifies this object as an eclipsing binary, which aligns better with an EW (W UMa-type contact binary) classification. The catalog period differs from that in Gaia (0.415448 days), likely contributing to the misclassification as an RRC.
\end{itemize}

\textbf{In-class Outliers}. Figure \ref{fig:inclass} displays objects that were flagged as outliers despite having correct labels. These stars were marked as outliers due to distinctive features:
\begin{itemize}
    \item EDR3 3017256242460492800, An EA-type star (Figure \ref{fig:inclass1}): identified as V1174 Ori, is a special X-ray bright pre-main sequence system in the Orion star-forming cluster \citep{2022ApJ...941..125S}.

    \item EDR3 3406832075676358912, Correctly classified as EB (Figure \ref{fig:inclass2}): shows unusual out-of-eclipse modulations, possibly from rotation.

    \item EDR3 3372147259826502272 (V* DU Gem), a semi-detached binary with emission lines (Figure \ref{fig:inclass3}).
    
    \item EDR3 45787237593398144, Both a misclassification and in-class outlier (Figure \ref{fig:inclass4}): likely an EB rather than EA, with a light curve suggesting rotation or pulsation effects.
\end{itemize}

\begin{table*}
\centering
\begin{tabular}{@{}l S[table-format=3.2] S[table-format=3.2] S[table-format=3.2] S[table-format=3.2] S[table-format=3.2] S[table-format=3.2] S[table-format=3.2]}
\toprule
\textbf{Class} & \textbf{Photometry} & \textbf{Spectra} & \textbf{Metadata} & \textbf{Photometry + Spectra} & \textbf{Spectra + Metadata} & \textbf{Metadata + Photometry} & \textbf{All} \\
\midrule
DSCT & 91.67 & 70.83 & 75.00 & 95.83 \greenuparrow & 75.00 & 95.83 \greenuparrow & 91.67\\
EA & 78.75 & 35.62 & 64.38 & 80.00 \greenuparrow & 65.62 \greenuparrow & 86.25 \greenuparrow & 85.62 \greenuparrow\\
EB & 80.00 & 58.75 & 47.50 & 76.25 \reddownarrow & 62.50 \greenuparrow & 76.88 \reddownarrow & 76.88 \reddownarrow\\
EW & 70.00 & 66.88 & 78.75 & 88.75 \greenuparrow & 76.88 \reddownarrow & 88.75 \greenuparrow & 90.00 \greenuparrow\\
HADS & 100.00 & 23.08 & 76.92 & 100.00 & 76.92 & 92.31 \reddownarrow & 96.15 \reddownarrow\\
M & 95.45 & 54.55 & 90.91 & 95.45 & 90.91 & 95.45 & 95.45\\
ROT & 91.25 & 84.38 & 92.50 & 97.50 \greenuparrow & 96.25 \greenuparrow & 99.38 \greenuparrow & 100.00 \greenuparrow\\
RRAB & 96.88 & 64.38 & 94.38 & 97.50 \greenuparrow & 95.00 \greenuparrow & 100.00 \greenuparrow & 100.00 \greenuparrow\\
RRC & 86.08 & 63.29 & 86.08 & 92.41 \greenuparrow & 86.08 & 94.94 \greenuparrow & 94.94 \greenuparrow\\
SR & 98.12 & 94.38 & 96.88 & 98.75 \greenuparrow & 98.12 \greenuparrow & 100.00 \greenuparrow & 100.00 \greenuparrow\\
\midrule
\textbf{Average} & 88.82 & 61.61 & 80.33 & 92.24 \greenuparrow & 82.33 \greenuparrow & 92.98 \greenuparrow & 93.07 \greenuparrow \\
\bottomrule
\end{tabular}
\caption{Class-wise accuracy for each modality (Photometry, Spectra, Metadata) and their combinations, showing the importance of each modality in the CLIP classification model. Green/red arrows indicate improved/decreased accuracy when modalities are combined. Results demonstrate that different classes benefit from different modalities, with all three modalities combined yielding the highest average accuracy across classes.}
\label{table:modality_contributions}
\end{table*}

\begin{figure}
    \centering
    \ifsvg
        \includesvg[width=0.75\columnwidth]{images/rot}
    \else
        \includegraphics[width=0.75\columnwidth]{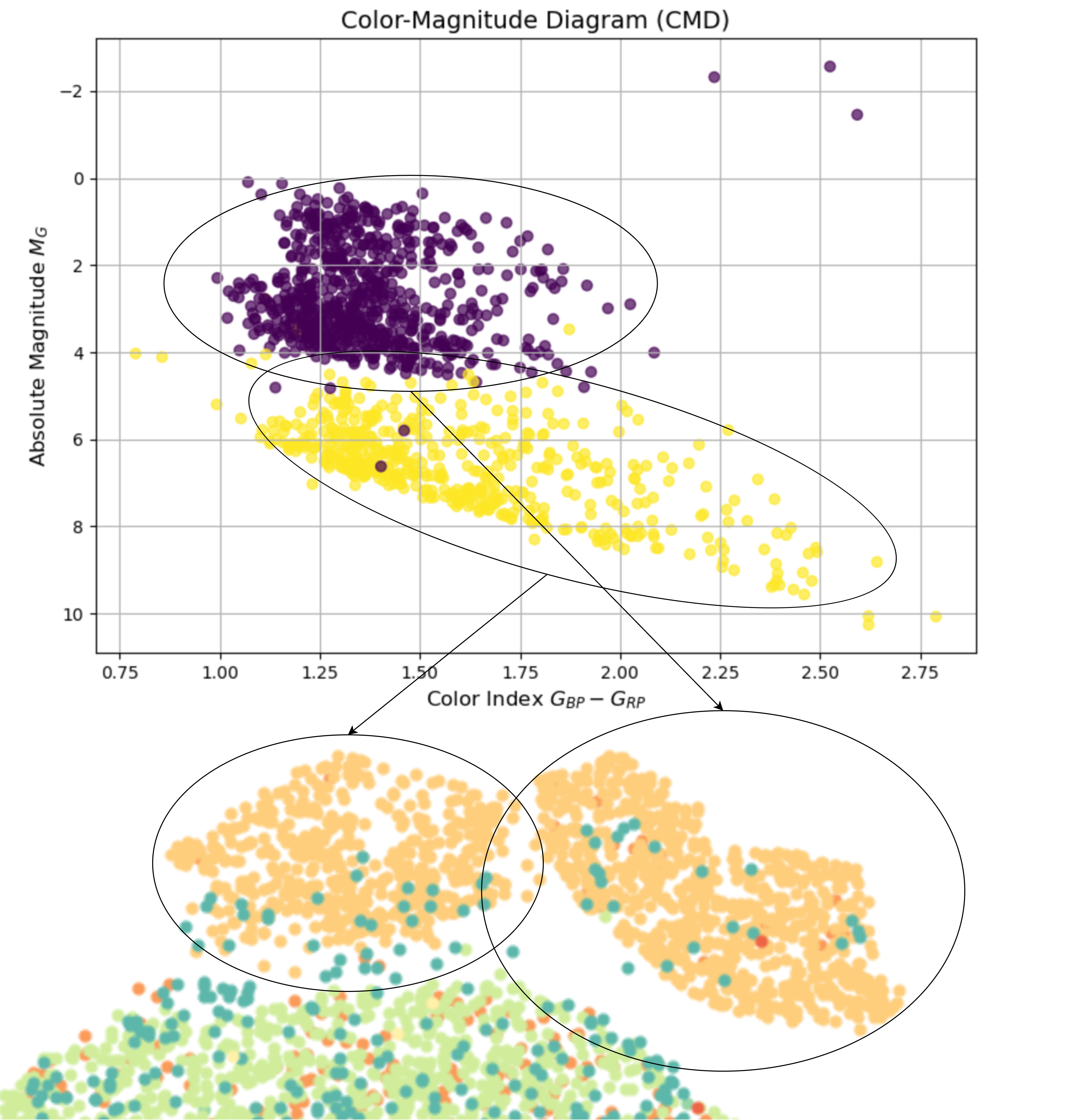}
    \fi
    \caption{Color-magnitude diagram for ROT variables, with two clusters identified through unsupervised learning as giants and dwarfs.}
    \label{fig:rot}
\end{figure}

\textbf{Two ROT Clusters.} Interestingly, the Spotted Variables with rotational modulation (ROT) class appears to be divided into two adjacent clusters, suggesting two physically distinct subtypes. To investigate further, we plotted these objects on a color-magnitude diagram (Figure \ref{fig:rot}). The plot revealed that the model had distinguished two subtypes within the ROT class: giants and dwarfs. Notably, the model discovered this distinction in an unsupervised learning process, without explicit labels for these subtypes.

\begin{figure}
    \centering
    \ifsvg
        \includesvg[width=\columnwidth]{images/mira}
    \else
        \includegraphics[width=\columnwidth]{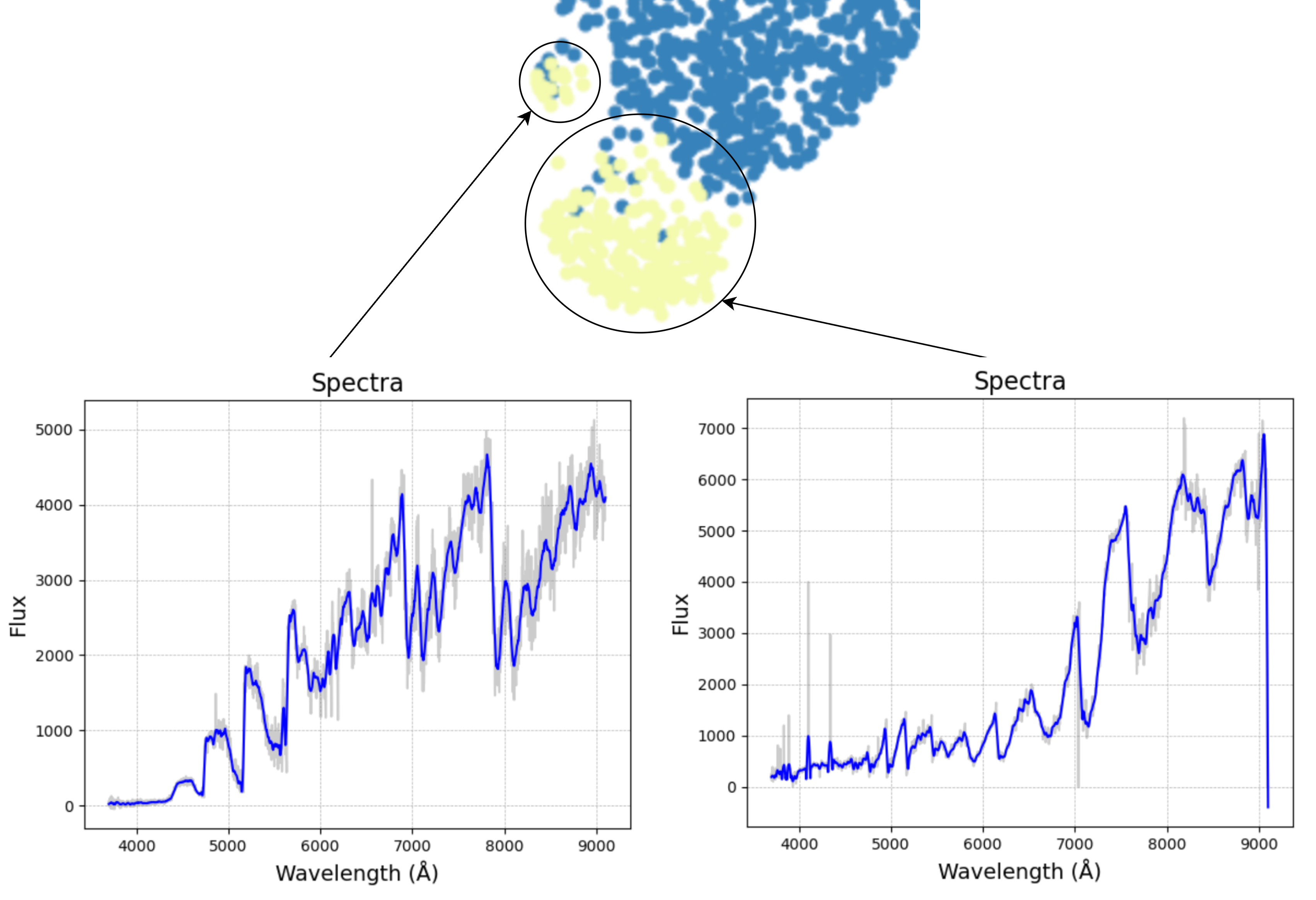}
    \fi
    \caption{Spectral examples of Mira variables, showing two distinct clusters corresponding to M-type and C-type Miras, discovered through unsupervised learning.}
    \label{fig:mira}
\end{figure}

\vspace{1em} 
\textbf{Two Mira Clusters.} Similarly in Figure \ref{fig:mira}, the Miras were also split into two clusters---one larger and one significantly smaller. Upon closer inspection, we find that these clusters correspond to two distinct subtypes of Miras: M-type and C-type. This distinction was not explicitly available beforehand, as our dataset only included the general "Mira" label. This demonstrates the ability of the approach taken herein to uncover hidden patterns in astronomical data and its potential for enabling new scientific discoveries.

\begin{figure}
    \centering
    \ifdownsampled
        \includegraphics[width=\columnwidth]{images/unseen-down.png} % Low-resolution version
    \else
        \includegraphics[width=\columnwidth]{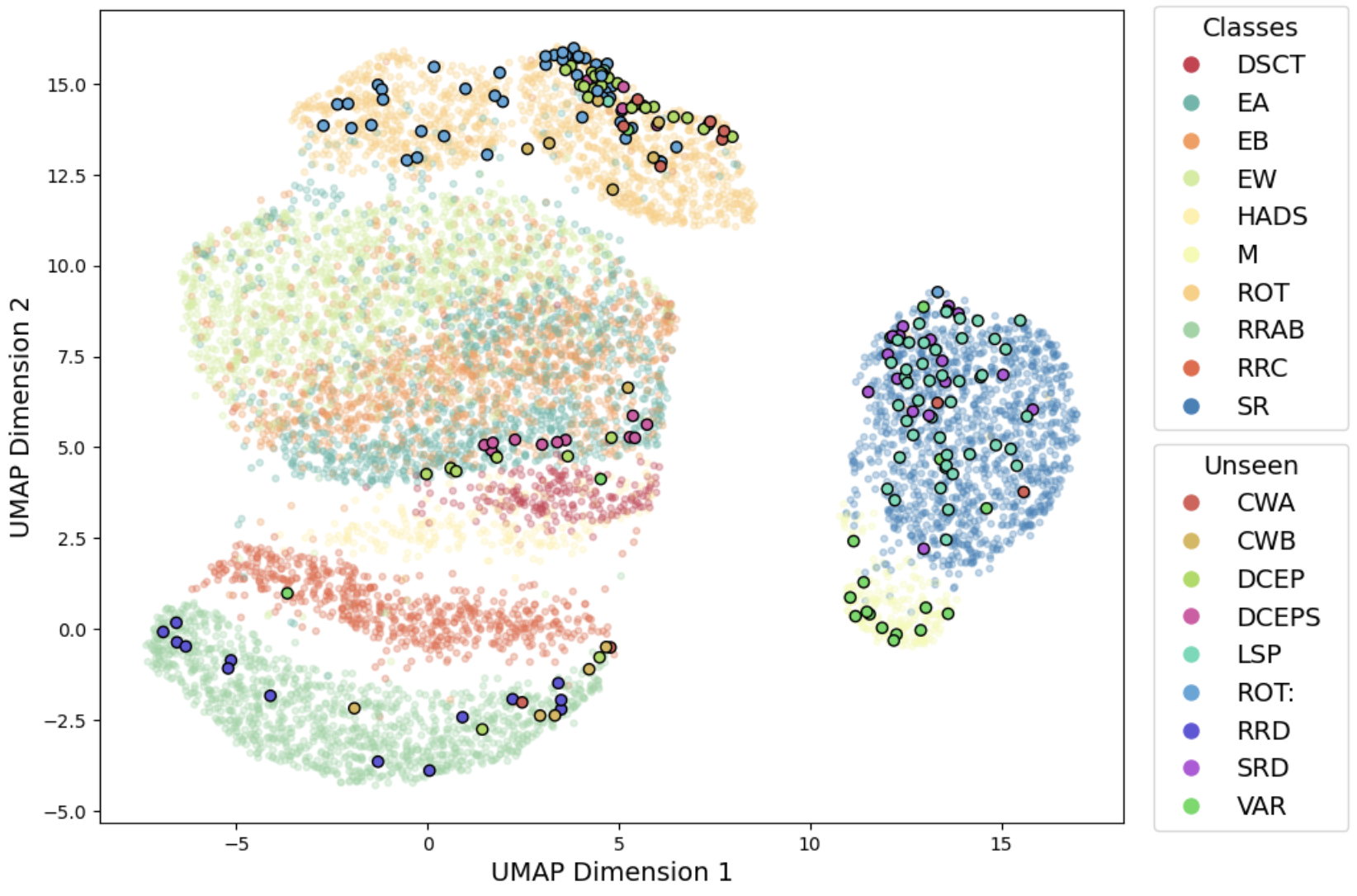} % Full-resolution version
    \fi
    \caption{Projections of new, previously unused classes in the embedding space, aligning with related clusters and demonstrating the model's ability to position unseen classes accurately}
    \label{fig:new_classes}
\end{figure}

\vspace{1em}
\textbf{New Classes.} During dataset creation, we filtered out classes with insufficient sample sizes. Now, with the learned embedding, we use these objects to test the ability of the model to project unseen classes. Figure \ref{fig:new_classes} shows they are located as expected: (a) Double Mode RR Lyrae variables (RRD) are located inside the cluster of RR Lyrae variables Type ab (RRAB); (b) uncertain Rotational variables (ROT:) within the certain ROT cluster; (c) Yellow semiregular variables (SRD) and Long Secondary Period (LSP) in the Semiregular variables (SR) cluster; (d) First overtone Cepheids (DCEPS) and some Fundamental mode Classical Cepheids (DCEP) near $\delta$ Scuti variables (DSCT). Interestingly, most uncertain classifications (VAR) fall within the Mira cluster.

\subsection{Modalities Importance}
To evaluate the importance of each modality in our CLIP classification model, we exploit the ability to utilize any combination of available modalities during testing. This flexibility is achieved by averaging the embeddings before the fully connected layer—rather than concatenating them—and by learning a shared embedding space. We calculate the class-wise accuracy percentages for each modality individually, for every pairwise combination, and for all modalities combined.

The results, presented in Table \ref{table:modality_contributions}, indicate that different modalities are crucial for different classes. For instance, the photometry modality is most significant for classes like DSCT, EA and EB, while metadata is more important for EW. Other classes benefit from more than one modality: ROT and RRAB show improved performance with both photometry and metadata, while SR achieves good accuracy with all three modalities.

Although the spectra modality alone yields lower accuracy than photometry, combining spectra with photometry results in equal or improved accuracy across all classes except for EB. The combination of spectra and metadata shows a similar pattern, achieving higher accuracy for all classes except EW. Likewise, combining metadata with photometry leads to equal or improved accuracy across all classes, with the exceptions of EB and HADS. On average, integrating any two modalities performs better than using a single modality, and combining all three modalities yields the highest accuracy overall.

\subsection{Similarity Search}

\begin{figure}
    \centering
    \ifsvg
        \includesvg[width=0.98\columnwidth]{images/similarity_search}
    \else
        \includegraphics[width=0.95\columnwidth]{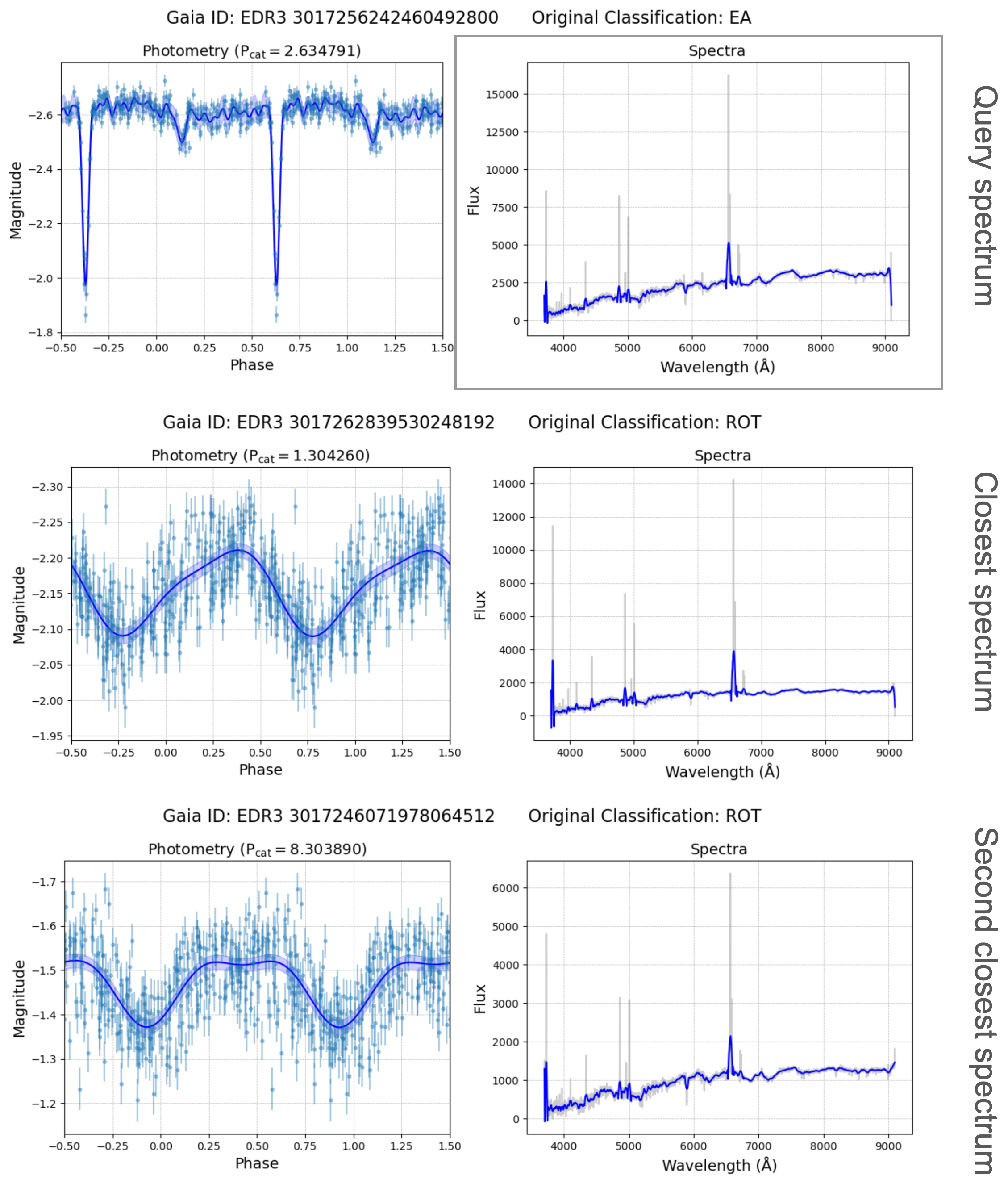}
    \fi
    \caption{The query spectrum (top) and its two closest matches (middle and bottom) based on spectral cosine similarity.}
    \label{fig:similar}
\end{figure}

An additional strength of our approach is the ability to perform similarity or dissimilarity searches within the embedding space. This expands the utility of the CLIP-based model beyond classification to serve as a versatile tool for exploratory data analysis, anomaly detection, and multimodal inference. This capability holds promise for aiding the discovery of rare or unexpected phenomena in astronomical data.

\vspace{1em}
\textbf{Modality-Specific Similarity Search.} Our model allows to find similar objects based on a chosen modality. For example, if we want to find objects with spectral features similar to those in Figure~\ref{fig:inclass1}, we can embed the spectrum of that object and compute the cosine similarity with other objects in the dataset (where a cosine similarity of 1 indicates maximum similarity). Figure~\ref{fig:similar} shows the two most similar objects based solely on spectral similarity, with cosine similarities of 0.8784 and 0.8451, respectively. As shown, they share clear visual similarities.

\begin{figure}
    \centering
    \ifsvg
        \includesvg[width=0.98\columnwidth]{images/contrast}
    \else
        \includegraphics[width=0.95\columnwidth]{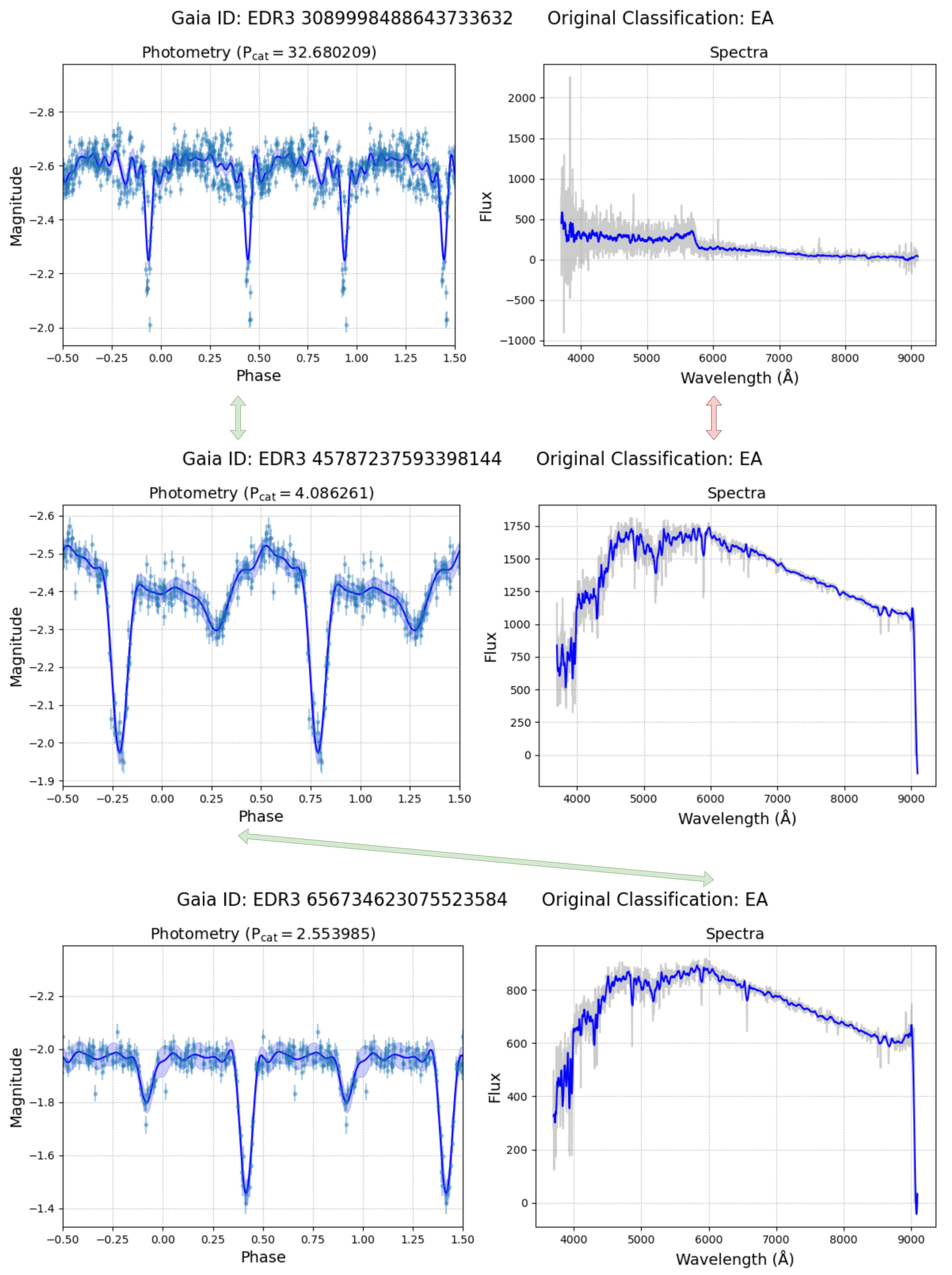}
    \fi
    \caption{Examples of cross-modality contrast and similarity searches. The object in the middle row serves as the query. The object in the top row has photometry similar to the query but shows distinct spectral characteristics. The object in the bottom row is identified by comparing the cosine similarity between its spectra and the photometry of the query object.}
    \label{fig:cross-mod}
\end{figure}

\vspace{1em}
\textbf{Cross-Modality Contrast Search.} Our approach also allows for searches to find objects that are similar in one modality but differ in another. For instance, we can first identify the 10 objects most similar to Figure~\ref{fig:inclass4} based on photometric cosine similarity. Among these, we then select the object with the greatest spectral difference. This process results in the object shown in Figure~\ref{fig:cross-mod}, which has a photometric cosine similarity of 0.7749 but a much lower spectral similarity of 0.1430. Notably, this object is also a misclassification with an incorrect period; the correct classification should be an RS Canum Venaticorum variable, with the actual period being half the reported value (16.3401046 days).

\vspace{1em}
\textbf{Cross-Modality Similarity Search.} When only photometric data is available, we can identify the closest matching spectra by calculating the cosine similarity between the photometric embedding and all the spectra embeddings in the dataset. This approach is possible because the model is trained to align photometry, spectra, and metadata in the same shared embedding space. For instance, using the photometry of the object shown in Figure \ref{fig:inclass4}, we find that the closest spectra in the dataset, as shown in Figure \ref{fig:cross-mod}, has a cosine similarity of 0.4872. Although there is no guarantee that the predicted spectra will perfectly match the actual spectra—especially given the relatively moderate cosine similarity—this method allows us to form hypotheses about an object's composition without requiring direct spectroscopic data.

\vspace{1em}
\textbf{Outlier Detection.} Beyond UMAP-based analysis, we can identify outliers using all 512 features of the embedding space. This allows us to detect (1) misclassifications, (2) in-class outliers, and (3) complete outliers that do not belong to any known class. To identify (1) and (2), we can calculate class centroids by averaging all embeddings for each class. We then build a cosine distance distribution for each class and set a threshold, such as the 99th percentile. Any object with a cosine distance from its class centroid exceeding this threshold can be labeled as an outlier. This process can be performed separately for each modality, and the results can be further refined by marking only those objects that are identified as outliers in more than one modality. For (3), we can apply DBSCAN clustering on the entire set of embeddings without using explicit labels, marking any object that falls outside the main clusters as a complete outlier.

\section{Conclusion}

We present the curation of a large labeled dataset suitable for building and testing next-generation multi-modal self-supervised models. This includes 21,440 objects with time-series photometry, spectra, and metadata. We also introduce AstroM$^3$ self-supervised pre-training framework that leverages all three data modalities. By extending the Contrastive Language-Image Pretraining model to handle a trimodal setting, our approach effectively learns joint representations across diverse astronomical data types, enhances classification accuracy, and leverages unlabeled data to improve performance when labeled data is limited. Beyond classification, AstroM$^3$ demonstrates versatility in tasks such as misclassification detection and in-class outlier identification. Additionally, it shows promise for scientific discovery by "rediscovering" different Mira types and Rotational variables subclasses, and enables efficient searches by identifying similar objects, cross-modality contrasts, or cross-modality similarities—facilitating targeted exploration of specific sources.

\vspace{1em}
\textbf{Future Work.} To be clear, while our approach outperforms classification tasks on the dataset we have curated, we are not claiming that AstroM$^3$ has been shown to achieve state-of-the-art on  classification of time-variable sources in general---the application of AstroM$^3$ to existing photometric benchmark datasets from other surveys is a clear next step. There are several other directions for extending our framework beyond AstroM$^3$. Given the abundance of photometry and metadata compared to spectra, one key area is to develop an algorithm capable of handling missing modalities {\it during training}, allowing us to leverage all available photometry and metadata. Additional directions include expanding the framework to integrate even more modalities, such as photometry from other bands and human comments on sources; learning to manage varying and missing metadata; and incorporating new classes, including non-periodic ones. Building a larger, more diverse dataset and applying the models to tasks like prediction and anomaly detection are essential next steps toward creating a truly foundational multimodal model for astronomy.

%%%%%%%%%%%%%%%%%%%%%%%%%%%%%%%%%%%%%%%%%%%%%%%%%%
\section*{Data Availability}

All code, model weights, parameters, and data will be made available upon acceptance. This includes the code for defining the models, as well as the code used for training and evaluation, the hyperparameters applied during training, the trained model weights, and the code for dataset construction, including all preprocessing pipelines. Although the raw data we used is publicly available, cross-matching and constructing the dataset required considerable effort, so we will also release the finalized, cross-matched dataset. Additionally, all software used in this work is open-source.

\section*{Acknowledgments}

M.\ R.\ and J.\ S. B.\ were partially supported by the Gordon and Betty Moore Foundation and an AAG grant  (\#2206744) from the National Science Foundation. J.\ S. B.\ thanks the Miller Institute for Basic Research in Science for supporting his research. The spectra used in this paper were from LAMOST (the Large Sky Area Multi-Object Fiber Spectroscopic Telescope) on the Guoshoujing Telescope, a  Major National Scientific Project built by the Chinese Academy of Sciences. Funding for the LAMOST project has been provided by the National Development and Reform Commission. LAMOST is operated and managed by the National Astronomical Observatories, Chinese Academy of Sciences. Photometry used in this paper were from ASAS-SN, funded in part by the Gordon and Betty Moore Foundation and also funded in part by the Alfred P. Sloan Foundation grant. This work also made use of data from the European Space Agency (ESA) mission Gaia (https://www.cosmos.esa.int/gaia), processed by the Gaia Data Processing and Analysis Consortium (DPAC, https: //www.cosmos.esa.int/web/gaia/dpac/consortium). Funding for the DPAC has been provided by national institutions, in particular the institutions participating in the Gaia Multilateral Agreement.



% \clearpage
\bibliographystyle{mnras}
\bibliography{paper} % if your bibtex file is called example.bib

%\begin{thebibliography}{99}
%\end{thebibliography}

%%%%%%%%%%%%%%%%%%%%%%%%%%%%%%%%%%%%%%%%%%%%%%%%%%

%%%%%%%%%%%%%%%%% APPENDICES %%%%%%%%%%%%%%%%%%%%%

% \clearpage
\appendix

\section{Training Setup and Hyperparameters}
\label{sec:hyperparameters}

In this work, we used Optuna \citep{akiba2019optuna} to perform hyperparameter optimization for our models. Our goal was to minimize the validation loss across multiple architectures and pre-training strategies. We tuned CLIP itself, as well as models for photometry, spectra, metadata, and multimodal data, with two initialization options: random initialization or pre-trained CLIP weights. 

For each model type, the hyperparameters we explored included:

\begin{itemize}
    \item Learning rate (\texttt{lr}): Sampled from a logarithmic scale between $1 \times 10^{-5}$ and $1 \times 10^{-2}$
    \item Dropout rates for photometry (\texttt{p\_dropout}), spectra (\texttt{s\_dropout}) and metadata (\texttt{m\_dropout}): All sampled from a uniform distribution between $0.0$ and $0.4$.
    \item Adam optimizer parameters:
    \begin{itemize}
        \item Beta1 (\texttt{beta1}): Sampled from a uniform distribution between $0.7$ and $0.99$.
        \item Weight decay (\texttt{weight\_decay}): Sampled from a logarithmic scale between $1 \times 10^{-5}$ and $1 \times 10^{-1}$.
    \end{itemize}
    \item Learning rate scheduler factor (\texttt{factor}): Sampled from a uniform distribution between $0.1$ and $1.0$ for the \texttt{ReduceLROnPlateau} scheduler.
\end{itemize}

\textbf{Training Setup.} For each trial, additional techniques were applied to ensure model stability and improve convergence:

\begin{itemize}
    \item Gradient clipping was applied to stabilize training. For CLIP, a clipping value of 45 was used, while for the photometry and spectra models, the clipping value was set to 5.
    \item Training duration: The models were trained for a fixed number of epochs: 100 epochs for CLIP and 50 epoch for others
    \item A warmup scheduler was employed to gradually increase the learning rate from a very low value to the target learning rate over the first 10 epochs.
    \item Early stopping based on validation loss was used with a patience of 6 epochs.
\end{itemize}

\begin{table}
\centering
\begin{tabular}{@{}ll@{}}
\toprule
\textbf{Feature} & \textbf{Description} \\ 
\midrule
mean\_vmag & Mean magnitude in the visible band \\ 
phot\_g\_mean\_mag & Gaia G-band mean magnitude \\ 
e\_phot\_g\_mean\_mag & Uncertainty in Gaia G-band mean magnitude \\ 
phot\_bp\_mean\_mag & Gaia BP band mean magnitude \\ 
e\_phot\_bp\_mean\_mag & Uncertainty in Gaia BP band mean magnitude \\ 
phot\_rp\_mean\_mag & Gaia RP band mean magnitude \\ 
e\_phot\_rp\_mean\_mag & Uncertainty in Gaia RP band mean magnitude \\ 
bp\_rp & BP mean magnitude minus RP mean magnitude \\ 
parallax & Gaia DR3 Parallax measurement \\ 
parallax\_error & Uncertainty in parallax measurement \\ 
parallax\_over\_error & Signal-to-noise ratio for parallax measurement \\ 
pmra & Proper motion in the Right Ascension direction\\ 
pmra\_error & Uncertainty in pmra \\ 
pmdec & Proper motion in the Declination direction\\ 
pmdec\_error & Uncertainty in pmdec \\ 
j\_mag & 2MASS J-band magnitude \\ 
e\_j\_mag & Uncertainty in 2MASS J-band magnitude \\ 
h\_mag & 2MASS H-band magnitude \\ 
e\_h\_mag & Uncertainty in 2MASS H-band magnitude \\ 
k\_mag & 2MASS K-band magnitude \\ 
e\_k\_mag & Uncertainty in 2MASS K-band magnitude \\ 
w1\_mag & WISE W1 band magnitude \\ 
e\_w1\_mag & Uncertainty in WISE W1 band magnitude \\ 
w2\_mag & WISE W2 band magnitude \\ 
e\_w2\_mag & Uncertainty in WISE W2 band magnitude \\ 
w3\_mag & WISE W3 band magnitude \\ 
w4\_mag & WISE W4 band magnitude \\ 
j\_k & J-band minus K-band magnitude \\ 
w1\_w2 & W1 band minus W2 band magnitude \\ 
w3\_w4 & W3 band minus W4 band magnitude \\ 
pm & Total proper motion \\ 
ruwe & Renormalized unit weight error \\ 
l & Galactic longitude \\ 
b & Galactic latitude \\ 
\bottomrule
\end{tabular}
\caption{Descriptions of metadata features used in the dataset.}
\label{table:feature_descriptions}
\end{table}

%%%%%%%%%%%%%%%%%%%%%%%%%%%%%%%%%%%%%%%%%%%%%%%%%%

% Don't change these lines
\bsp	% typesetting comment
\label{lastpage}
\end{document}